\documentclass[11pt]{article}
\usepackage{latexsym}
\usepackage{amssymb}
\usepackage{amsmath}
\usepackage{mathrsfs}
\usepackage{psfrag}
\usepackage{graphicx}
\usepackage{epstopdf}
	\DeclareGraphicsExtensions{.eps}
\usepackage{multicol}
\usepackage{caption}
\usepackage{amssymb}
\newtheorem{thm}{Theorem}[section]

\newtheorem{prop}[thm]{Proposition}
\newtheorem{defn}[thm]{Definition}

\newtheorem{exam}[thm]{Example}

\begin{document}

\vspace*{.5cm}
\def\cl{\centerline}

\begin{center}
{\LARGE{\textbf{On relative weighted entropies with central moments weight functions}}}
\end{center}
\bigskip
\cl{\Large{\bf Salimeh Yasaei Sekeh$^\dag$, Adriano Polpo $^*$}}
\footnote{
{\it $^\dag$ Department of Statistics, Federal University of S$\tilde{a}$o Carlos (UFSCar), BRAZIL}\\
\it $^\dag$ E-mail: sa$_{-}$yasaei@yahoo.com\\
\it  $^*$ E-mail: polpo@ufscar.br}


\bigskip


\vspace{0.5cm}
\begin{abstract}
Following \cite{belis1968}, the aim of this paper is to analyze the relative weighted entropy involving the central moments weight functions. We compare the standard relative entropy with the weighted case in two particular forms of Gaussian distributions. As an application, the weighted deviance information criterion is proposed.

\end{abstract}


\def\fB{\mathfrak B}\def\fM{\mathfrak M}\def\fX{\mathfrak X}
 \def\cB{\mathcal B}\def\cM{\mathcal M}\def\cX{\mathcal X}
\def\bu{\mathbf u}\def\bv{\mathbf v}\def\bx{\mathbf x}\def\by{\mathbf y}
\def\om{\omega} \def\Om{\Omega}
\def\bbP{\mathbb P} \def\hw{h^{\rm w}} \def\hwi{{h^{\rm w}}}
\def\beq{\begin{eqnarray}} \def\eeq{\end{eqnarray}}
\def\beqq{\begin{eqnarray*}} \def\eeqq{\end{eqnarray*}}
\def\rd{{\rm d}} \def\Dwphi{{D^{\rm w}_\phi}}
\def\BX{\mathbf{X}}\def\Lam{\Lambda}\def\BY{\mathbf{Y}}

\def\mwe{{D^{\rm w}_\phi}}
\def\DwPhi{{D^{\rm w}_\Phi}} \def\iw{i^{\rm w}_{\phi}}
\def\bE{\mathbb{E}}
\def\1{{\mathbf 1}} \def\fB{{\mathfrak B}}  \def\fM{{\mathfrak M}}
\def\diy{\displaystyle} \def\bbE{{\mathbb E}} \def\bu{\mathbf u}
\def\BC{{\mathbf C}} \def\lam{\lambda} \def\bbB{{\mathbb B}}
\def\bbR{{\mathbb R}}\def\bbS{{\mathbb S}} \def\bmu{{\mbox{\boldmath${\mu}$}}}
 \def\bPhi{{\mbox{\boldmath${\Phi}$}}}
 \def\bbZ{{\mathbb Z}} \def\fF{\mathfrak F}\def\bt{\mathbf t}\def\B1{\mathbf 1}
\def\UX{\underline{X}}
\def\ux{\underline{x}}
\def\Hw{H^{\rm w}}
\def\OY{\overline{Y}} \def\oy{\overline{y}}
\def\OY{\overline{Y}} \def\oy{\overline{y}} \def\omu{\overline{\mu}} \def\OSigma{\overline{\Sigma}}

\section{Introduction: The weighted entropies}
Let $\UX$ be a real-valued random vector (RV) with a join probability density function (PDF) $f$. The differential entropy (DE) of
RV $\underline{X}$ is defined by
\beq \label{eq:DE0}H(\UX)=H(f)=\diy\int _{\bbR^n} f(\ux)\log\;f(\ux)\rd \ux. \eeq

The definition and a number of inequalities for a standard DE were illustrated in \cite{shannon1948, cover2006, KS}. Furthermore, in \cite{belis1968,Guiasu1986} the initial definition and results on weighted entropy was introduced. Following \cite{ShMM1978, DT1981, PT1986, SiB1992}, recently in \cite{di crescenzo2006, Clim2008, S2011, suhov2014, suhov2015}, a similar method with standard DE drives to emerge certain properties and applications of information-theoretical weighted entropies with a number of determinant-related inequalities.

Let $\ux\in\bbR^n\mapsto\phi (\ux) \geq 0$ be a given (measurable) function, called weighted function (WF). The weighted differential entropy (WDE) $H^{\rm w}_\phi (\UX)$ of a real-valued RV $\UX$
with a PDF $f$ is given by
\beq\label{eq:WDE0}H^{\rm w}_\phi (\UX) =H^{\rm w}_\phi (f):=-\bbE\,\phi (\UX)\log\,f(\UX) =-\int_{\bbR^n}\phi (\ux )f(\ux)\log\,f(\ux )\rd \ux ,\eeq
Note that the WDE (\ref{eq:WDE0}) is obtained for a given non-negative WF; when this function equals 1, the WDE coincides with the standard (Shannon) DE, (\ref{eq:DE0}).\\
We also assume the integrals in (\ref{eq:DE0}) and (\ref{eq:WDE0}) absolutely converge, on the other hand the WDE and DE are finite.
A standard agreement $0=0.\log 0=0. \log\infty$ is considered throughout the paper.

We now give the definition of conditional DE and mutual DE for RVs, in
view of the fact that these are the ones on which we shall focus in our analysis more.
\begin{defn}\label{defn mutual}
Let $\UX_1\in\bbR^{m_1}$ and $\UX_2\in\bbR^{m_2}$ be two RVs, with joint PDFs $f(\ux_1,\ux_2)$ and marginal PDFs $f_1(\ux_1)$ and $f_2(\ux_2)$. The conditional DE of $\UX_1$ given $\UX_2$ is defined by
\beq \label{con.DE} H(\UX_1|\UX_2)=-\int_{\bbR^{m_1+m_2}} f(\ux_1,\ux_2)\log \frac{f(\ux_1,\ux_2)}
{f_2(\ux_2)}\rd \ux_1 \rd\ux_2.\eeq
Next for RV $\UX=(X_1,X_2,\dots,X_n)$, we use joint and marginal PDFs $f_{X_1,\dots,X_n}$ and $f_{X_1},f_{X_2},\dots,f_{X_n}$
to define the mutual DE by
\begin{eqnarray}\label{mutual inform.}
I(f_{X_1,\dots,X_n}, f_{X_1}\dots f_{X_n})=\int_{\mathbb{R}^n}f(\ux) \log \frac{f(\ux)}{f_1(x_1) \dots f_n(x_n)} d\underline{x},
\end{eqnarray}
note that motivated by continuity, we set $0\log \frac{0}{0}=0$.
\end{defn}
Here and below we use both notations $f(x_1,\dots,x_n)$ and $f_{X_1,\dots,X_n}$ for joint PDF allowing us to be flexible in shortening throughout the paper. In addition we employ both $f_i(x_i)$ and $f_{X_i}$ as marginal PDF of random variable $X_i$, $i=1,\dots,n$.\\

The following theorem was proven in \cite{cover2006}.
\begin{thm} {\rm{(Chain rule for the DE)}} Let $X_1,\dots, X_n$ be drown according to joint density PDF $f(x_1,\dots,x_n)$, then
\begin{eqnarray}
H(X_1,X_2,\dots,X_n)=\sum_{i=1}^n H(X_i|X_{i-1},\dots,X_1).
\end{eqnarray}
\end{thm}

One of the mutual information's properties, the same as bivariate case, is which can be demonstrated also in terms of
marginal entropy and conditional entropy. The proof comes directly if we rewrite (\ref{mutual inform.}) in Definition \ref{defn mutual} and omitted.
\begin{prop}
For RV $\UX\in\bbR^n$ with joint and marginal PDFs $f_{X_1,\dots,X_n}$ and $f_{X_1},f_{X_2},\dots,f_{X_n}$, we have
\begin{eqnarray}
I(f_{X_1,\dots,X_n}, f_{X_1}\dots f_{X_n})=\sum_{i=1}^{n-1} \left[H(X_i)-H(X_i|X_{i+1},\dots,X_n)\right].
\end{eqnarray}
\end{prop}
\vskip 1 truecm
{\bf Remark:} An alternative expression for mutual information, in terms of entropy and conditional entropy is derived as follows:
\begin{eqnarray}
I(f_{X_1,\dots,X_n}, f_{X_1}\dots f_{X_n})=\sum_{i=1}^{n-1} E_{X_{i+1},\dots,X_n}\left[ H(X_i)-H(X_i|x_{i+1},\dots,x_n)\right].
\end{eqnarray}
Here
\beq\label{DE:givenRV} H(X_i|x_{i+1},\dots,x_n)=-\int_{\bbR}f(x_i|x_{i+1},\dots,x_n)\log f(x_i|x_{i+1},\dots,x_n)\rd x_i.\eeq

As the Definition 1.2 in \cite{suhov2014}: Let $\ux\in\bbR^n\mapsto\phi (\ux) \geq 0$ be a WF. The conditional WE of RV $\UX_1\in\bbR^{m_1}$
 given $\UX_2\in\bbR^{m_2}$ is defined by
\beq H^{\rm w}_\phi(\UX_1|\UX_2)=-\int_{\bbR^{m_1+m_2}} \phi(\ux_1,\ux_2) f(\ux_1,\ux_2)
\log \frac{f(\ux_1,\ux_2)}{f_2(\ux_2)}\rd \ux_1 \rd\ux_2,\eeq
and the mutual WE is given by
\beq \label{WMI}I^{\rm w}_\phi(f_{X_1,\dots,X_n}, f_{X_1}\dots f_{X_n})=\int_{\mathbb{R}^n} \phi(\ux) f(\ux)
 \log \frac{f(\ux)}{f_1(x_1) \dots f_n(x_n)} d\underline{x}.\eeq
This concept is easily adapted to the weighted DE by using the quality of random variables, as explained in \cite{belis1968}.

\section{Relative weighted entropies}
The contribution of our paper in this setting is thus twofold:
\begin{enumerate}
  \item We briefly improve several theorems discovered in \cite{yasaei2014} and give alternative definitions in particular form of WF.
  \item We reformulate these results for Gaussian distribution with two different covariance matrixes.
\end{enumerate}
\subsection{Central moments weight functions}

As we said in this paper, basically this subsection, we deal with central moments WFs, of the form  $\phi(\ux)=\prod\limits_{i=1}^n\; (x_i-a_i)^2$ for constants $a_1,\dots,a_n$.


The naturalness of the definition of the WE and conditional WE is exhibited by the fact that the WE of a vector of random variables is
the conditional WE of one plus the generalized conditional WE of the others. On the other hand the chain rule can thus be
adapted to the WE; accordingly, we reformulate it as follows:
\begin{thm}{\rm{(Chain rule for the WE)}} Consider the RV $\underline{X}=(X_1,\dots,X_n)$ with joint PDF $f(x_1,\dots,x_n)$. Then for constants
$a_1,\dots,a_n$ and given WF $\phi(\ux)=\prod\limits_{i=1}^n (x_i-a_i)^2$
\begin{eqnarray*}
H^{\rm w}_\phi(X_1,\dots,X_n)=H^{\rm w}_\phi(X_n|X_{n-1},\dots,X_1)+\sum_{i=1}^{n-1}H^{\rm w}_{\psi_i}(X_i|X_{i-1},\dots,X_1).
\end{eqnarray*}
Here for constants $a_1,\dots,a_n$
\beq\psi_i(x_1,\dots,x_i)=\prod_{j=1}^i (x_j-a_j)^2 \ E\bigg((X_{i+1}-a_{i+1})^2|(X_i,\dots,X_1)=(x_i,\dots,x_1)\bigg)\eeq.
\end{thm}

{\bf Proof}: If $(X_1,X_2)$ is a random pair, then in this particular case we have,
\beqq H^{\rm w}_\phi(X_1,X_2)=\Hw_\phi(X_2|X_1)+\Hw_{\psi_1}(X_1),\eeqq
Note that here $\psi_1(x_1)=(x_1-a_1)^2\;E((X_2-a_2)^2|X_1=x_1)$. Now more generally assume triple random $(X_1,X_2,X_3)$, similarly the WE is obtained by
\beqq\Hw_\phi(X_1,X_2,X_3)=\Hw_\phi(X_3|X_2,X_1)+\Hw_{\psi_2}(X_2|X_1)+\Hw_{\psi_1}(X_1),\eeqq
The WFs $\psi_1$ and $\psi_2$ are given by using the form of $\psi_i$ when $i=1,2$.
Applying the same methodology and expanding the RV to $n$ random variables, $n>3$ we detect the given form by
\begin{eqnarray*}\begin{array}{l}
\Hw_\phi(X_1,\dots,X_n)=\Hw_\phi(X_2,\dots,X_n|X_1)+\Hw_{\psi_1}(X_1)\\
\quad=\diy \Hw_\phi(X_3,\dots,X_n|X_2,X_1)+\Hw_{\psi_2}(X_2|X_1)+\Hw_{\psi_1}(X_1)\\
\quad...\\
\quad=\diy\Hw_\phi(X_n|X_{n-1},\dots,X_1)+\Hw_{\psi_{n-1}}(X_{n-1}|X_{n-2},\dots,X_1)+\\
\quad...\diy+\Hw_{\psi_{n-1}}(X_{n-2}|X_{n-3},\dots,X_1)+\Hw_{\psi_2}(X_2|X_1)+\Hw_{\psi_1}(X_1).\end{array}
\end{eqnarray*}
This leads to the desired result. $\quad$ $\Box$

In this stage, an immediate question crossed our mind which states shall we extend the similar
conclusions due to the weighted entropies? 
In fact, among all equivalent expression for the mutual WE, as already observed in mutual information,
 the most applicable is represented by the WE and the conditional WE.
\begin{thm}
Let us now consider the weighted mutual information, $I^{\rm w}_\phi(f_{X_1,\dots,X_n}, f_{X_1}\dots f_{X_n})$, then it can be written as follows:
\begin{eqnarray}
I^{\rm w}_\phi(f_{X_1,\dots,X_n}, f_{X_1}\dots f_{X_n})=\sum_{j=1}^{n-1} \Hw_{\psi'_j}(X_j)-\Hw_\phi(X_1,\dots,X_{n-1}|X_n),
\end{eqnarray}
where $\psi'_j(x_j)=(x_j-a_j)^2\;E\left[\prod_{i=1,i\neq j}^{n}(X_i-a_i)^2|X_j=x_j\right]$.
\end{thm}

{\bf Proof}: By recalling (\ref{WMI}), we observe that
\begin{eqnarray*}\begin{array}{l}
I^{\rm w}_\phi(f_{X_1,\dots,X_n}, f_{X_1}\dots f_{X_n})\\
\qquad= \diy \int_{\mathbb{R}^n}\prod_{i=1}^n (x_i-a_i)^2 \; f(x_1,\dots,x_n)
 \log f(x_1,\dots,x_{n-1}|x_n) \rd\underline{x}\\
\qquad-\diy\int_{\mathbb{R}^n}\prod_{i=1}^n (x_i-a_i)^2 \; f(x_1,\dots,x_n) \sum_{j=1}^{n-1}\log f_j(x_j)\rd\underline{x}\end{array}
\end{eqnarray*}
Consequently,
\begin{eqnarray*}\begin{array}{l}
I^{\rm w}_\phi(f_{X_1,\dots,X_n}, f_{X_1}\dots f_{X_n})=-\Hw_\phi(X_1,\dots,X_{n-1}|X_n)\\
\quad-\diy\int_\mathbb{R}\sum_{j=1}^{n-1} (x_j-a_j)^2
 \int_{\mathbb{R}^{n-1}}\prod_{i=1,i\neq j}^{n} (x_i-a_i)^2 f(x_1\dots,x_{j-1},x_{j+1},\dots,x_n|x_j)f_j(x_j) \log f_j(x_j) \rd\underline{x}\\
\quad=\diy-\Hw_\phi(X_1,\dots,X_{n-1}|X_n)-\diy\sum_{j=1}^{n-1}\int_\mathbb{R}(x_j-a_j)^2 E\left[\prod_{i=1,i\neq j}^{n}(X_i-a_i)^2|X_j\right] f_j(x_j)
\log f_j(x_j) \rd x_j.\end{array}
\end{eqnarray*}
Which is precisely the result that we are looking for. $\quad$ $\Box$\\

Considering real situation which there exist two dependent groups of components or on the other hand random vectors,
 in some experimental research we are looking for the discrimination between probability function while such vectors
 are independent and dependent, in fact using this methodology clarifies the effect of dependent random vectors.
 Indeed, it seems as much as the dependency between two groups of random data is stronger than the information among
 density functions should raise.  This fact will be adopted specifically in Gaussian distribution in the next subsection throughout examples.
\begin{prop}\label{conditionalK-B}
Suppose that $\underline{X}=(X_1,\dots,X_n)$ and $\underline{Y}=(Y_1,\dots,Y_m)$ be RVs showing any real situation,
with joint PDF $f(x_1,\dots,x_n,y_1,\dots,y_m)$ and marginal multivariate PDFs $f_1(x_1,\dots,x_n)$ and $f_2(y_1,\dots,y_m)$ respectively. Then
\begin{eqnarray}\label{dpkli}
D(f_{\underline{X}|\underline{y}}\|f_{\underline{X}})=\Hw_{\phi_{\underline{X}|\underline{y}}}(\underline{X})-H(\underline{X}|\underline{y}),
\end{eqnarray}
here $\phi_{\underline{X}|\underline{y}}(\ux)=f(\underline{x}|\underline{y})\Big/f_1(\underline{x})$ and $H(\underline{X}|\underline{y})$ is defined as (\ref{DE:givenRV}).

\vskip .5 truecm

In addition, let us here define the relative DE, known as the Kullback-Leibler divergence, for two given functions $f$ and $g$, $D(f\|g)$ by
\beq\label{KLI} D(f\|g)=\diy\int_{\bbR^n} f(\ux)\log\frac{f(\ux)}{g(\ux)}\rd\ux.\eeq
\end{prop}

{\bf Proof}: The Proof is based on the equation (\ref{KLI}) and straightforward. $\quad$ $\Box$

\vskip .5 truecm

{\bf Remak:} We explicitly note that by taking expectation on $D(f_{\underline{X}|\underline{y}},f_{\underline{X}})$
 with respect to RV $\underline{Y}$, by virtue of (\ref{dpkli}), mutual DE can be yielded:
\begin{eqnarray}\label{IKLC}
I(f_{\underline{X},\underline{Y}},f_{\underline{X}}f_{\underline{Y}})=E_{\underline{Y}}\left[D(f_{\underline{X}|\underline{y}}
,f_{\underline{X}})\right]
=E_{\underline{Y}}\left[\Hw_{\phi_{\underline{X}|\underline{y}}}(\underline{X})-H(\underline{X}|\underline{y})\right].
\end{eqnarray}
\vskip .5 truecm

Here, going back to weighted information measure we implicitly present the weighted information between
$f_{\underline{X}\mid \underline{y}}$ and $f_{\underline{X}}$. This probably makes reader even more interested,
however we were also wondering whether the amount value of RVs associates the effects of dependency
or not but let us to concentrate on this object in the next part of the paper.

\begin{defn}
For two functions $\bx\in\bbR^n\mapsto f(\bx)\geq 0$ and $\bx\in\bbR^n\mapsto g(\bx)\geq 0$, the relative WE
(the weighted Kullback-Leibler divergence ), for given WF $\phi$ is defined by
\beq\label{WKLI} D^{\rm w}_\phi(f\|g)=\int_{\bbR^n} \phi(\ux) f(\ux)\log\;\frac{f(\ux)}{g(\ux)}\;\rd \ux.\eeq
\end{defn}
\vskip .5 truecm
\begin{thm}\label{cor.1}
With the same assumptions and analogue method as Proposition \ref{conditionalK-B}, for given WF $\phi(\ux)=\prod\limits_{i=1}^n (x_i-a_i)^2$,
 constants $a_1,\dots,a_n$, one can obtain the following relationship:
\begin{eqnarray}\label{conditionalWK-B}\begin{array}{l}
D^{\rm w}_{\phi}(f_{\underline{X}|\underline{y}},f_{\underline{X}})=\diy\int_{\mathbb{R}^n} \prod_{i=1}^n (x_i-a_i)^2
\ f(\underline{x}|\underline{y}) \log \frac{f(\underline{x}|\underline{y})}{f(\underline{x})}\rd\underline{x}\\
\quad=\diy\int_{\mathbb{R}^n} \prod_{i=1}^n (x_i-a_i)^2\; f(\underline{x}|\underline{y})\log f(\underline{x}|\underline{y})\rd\underline{x}\\
\qquad-\diy\int_{\mathbb{R}^n} \prod_{i=1}^n (x_i-a_i)^2\; \frac{f(\underline{x}|\underline{y})}{f(\underline{x})}
f(\underline{x})\log f(\underline{x}) \rd\underline{x}\\
\\
\quad=\Hw_{\phi'_{\underline{X}|\underline{y}}}(\underline{X})-\Hw_\phi(\underline{X}|\underline{y}).\end{array}
\end{eqnarray}
where
\beq\label{WFphi'}\phi'_{\underline{X}|\underline{y}}(\ux)=\prod_{i=1}^n (x_i-a_i)^2\left[ \frac{f(\underline{x}|\underline{y})}
{f_1(\underline{x})}\right]\eeq
\end{thm}
Similar equations to (\ref{IKLC}) in terms of weighted case also can be seen,
\begin{eqnarray*}\begin{array}{l}
E_{\underline{Y}}\left[\prod_{j=1}^m (Y_j-a_j)^2\; D^{\rm w}_\phi(f_{\underline{X}|\underline{y}},f_{\underline{X}})\right]\\
\qquad=\diy\int_{\mathbb{R}^m} \prod_{j=1}^m (y_j-a_j)^2 \ f_2(\underline{y}) \int_{\mathbb{R}^n} \prod_{i=1}^n (x_i-a_i)^2
\; f(\underline{x}|\underline{y})\log \frac{f(\underline{x}|\underline{y})}{f_1(\underline{x})} d\underline{x} \rd\underline{y}\\
\qquad=\diy\int_{\mathbb{R}^{n+m}} \prod_{j=1}^m (y_j-a_j)^2\prod_{i=1}^n (x_i-a_i)^2
\;f(\underline{x},\underline{y})\log \left[ \frac{f(\underline{x},\underline{y})}{f_1(\underline{x})f_2(\underline{y})} \right]
d\underline{x} \rd\underline{y},\end{array}
\end{eqnarray*}
hence at last piece of discussion in this subsection, we point out that the mutual WE can be implied by
calculating $m$th order of moments for random vector $\underline{Y}$ while $D^{\rm w}_\phi(f_{\underline{X}|\underline{y}},
f_{\underline{X}}). f(\underline{y})$ plays the rule of density function.
\begin{eqnarray}\begin{array}{l}
\diy I^{\rm w}_\phi(f_{\underline{X},\underline{Y}},f_{\underline{X}}f_{\underline{Y}})=\diy E_{\underline{Y}}\left[\prod_{j=1}^m (Y_j-a_j)^2
\; D^{\rm w}_\phi(f_{\underline{X}|\underline{y}},f_{\underline{X}})\right]\\
\qquad=\diy E_{\underline{Y}}\left[\prod_{j=1}^m (Y_j-a_j)^2
\; \left\{\Hw_{\phi'_{\underline{X}|\underline{y}}}(\underline{X})-\Hw_\phi(\underline{X}|\underline{y})\right\} \right].\end{array}
\end{eqnarray}
The WF $\phi'_{\underline{X}|\underline{y}}$ applies the form as in (\ref{WFphi'}).


\subsection{Gaussian distribution}

The Gaussian distribution is the most useful, and most studied, of the standard joint distributions in probability.
A huge body of statistical theory depends on the properties of families  of random variables whose joint distribution
 is at least approximately multivariate normal. As we know many fancy statistical procedures implicitly require bivariate
(or multivariate, for more than two random variables) normality. Moreover, the hypothesis of dependency between random
variables has been always the center of researcher's attentions, hence in this subsection we focus on the dependent
RVs with Gaussian distribution.\\

Throughout this part of our research we give two types of Gaussian examples with different covariance matrixes.
By using the same technique as before, general formulas for $n=3$ are given. Furthermore we will observe the rule of
 coefficient correlation $\rho$ in the relative measure for the weighted and standard forms.\\

Consider $\underline{X}\sim \mathcal{N}(\underline{\mu},\underline{\Sigma})$,
 one of the achievements in \cite{cover2006} explicitly shows that the entropy for this famous family does not depend on $\underline{\mu}$:
\begin{eqnarray}\label{multivariate entropy}
H(\underline{X})=\frac{1}{2} \log \left[(2\pi)^n |\Sigma|\right].
\end{eqnarray}
where $|\Sigma|$ is the determinant matrix $\Sigma$.\\

However, in the Gaussian case, by virtue of (\ref{eq:WDE0}) involving WF $\phi(\ux)=\prod\limits_{i=1}^3(x_i-\mu_i)^2$, the WE admits a representation depending on mean $\underline{\mu}$, (see the Appendix):
\begin{eqnarray}\label{weightednormal}
\Hw_{\phi}(\UX)=\diy\left[ \frac{1}{2} \log \left((2\pi)^3 |\Sigma|\right)\right]\;\Xi
+\frac{1}{2}\sum_{i=1}^3 \sum_{j=1}^3 \Sigma_{ij}^{-1}\Lambda_{ij}.
\end{eqnarray}
Here
\beq\label{Lambda.Xi}\begin{array}{l} \Lambda_{ij}=\diy \bigg(\Sigma_{11}\Sigma_{22}+2\big(\Sigma_{12}\big)^2\bigg)
.\bigg(\Sigma_{33}\Sigma_{ij}+2\;\Sigma_{3i}\Sigma_{3j}\bigg),\\
\Xi=\diy \Sigma_{11}\bigg(\Sigma_{22}\Sigma_{33}+2\;(\Sigma_{23})^2\bigg)+2\;\Sigma_{12}\bigg(\Sigma_{12}\Sigma_{33}+2\Sigma_{13}\Sigma_{23}\bigg)\\
\qquad +2\;\Sigma_{13}\bigg(2\Sigma_{12}\Sigma_{23}+\Sigma_{13}\Sigma_{22}\bigg).\end{array}\eeq
\vskip .5 truecm
Recall random pair $\underline{X}_1=(X_1,X_2)$, having Gaussian distribution with mean
$ \underline{\mu}_1=\left[
                      \begin{array}{c}
                        \mu_1 \\
                        \mu_2 \\
                      \end{array}
                    \right]$ and covariance matrix
$\Sigma_1=\left[
            \begin{array}{cc}
              \Sigma_{11} & \Sigma_{12} \\
              \Sigma_{21} & \Sigma_{22} \\
            \end{array}
          \right]$. Hence the RV $(X_1,X_2|X_3=x_3)\sim \mathcal{N}(\overline{\mu},\overline{\Sigma})$, where
\beq\label{Omu} \overline{\mu}= \left[
                    \begin{array}{c}
                      \overline{\mu}_1 \\
                      \\
                      \overline{\mu}_2 \\
                    \end{array}
                  \right]= \left[
                   \begin{array}{c}
                     \mu_1+\Sigma_{13}\Sigma_{33}^{-1}(x_3-\mu_3) \\
                     \\
                     \mu_2+\Sigma_{23}\Sigma_{33}^{-1}(x_3-\mu_3) \\
                   \end{array}
                 \right],\eeq
\beq\label{OSigma} \overline{\Sigma}=\left[\begin{array}{cc}
                           \overline{\Sigma}_{11} & \overline{\Sigma}_{12} \\
                           \overline{\Sigma}_{21} & \overline{\Sigma}_{22}
                         \end{array}\right]=\left[
                                             \begin{array}{cc}
                                               \Sigma_{11}-\Sigma_{13}\Sigma_{33}^{-1}\Sigma_{31} \ \ \ & \ \ \
                                                \Sigma_{12}-\Sigma_{13}\Sigma_{33}^{-1}\Sigma_{32} \\
                                               \\
                                               \Sigma_{21}-\Sigma_{23}\Sigma_{33}^{-1}\Sigma_{31} \ \ \ & \ \ \
                                                \Sigma_{22}-\Sigma_{23}\Sigma_{33}^{-1}\Sigma_{32} \\
                                             \end{array}
                                           \right].\eeq
Further, we represent the inclosed formula for $D(f_{(X_1,X_2)|X_3},f_{(X_1,X_2)})$. Then we shall exploit later to compare it with the weighted one in order to catch a part of our main purpose in this work.\\

Going back to the Proposition \ref{conditionalK-B} admits the representation
\begin{eqnarray}\label{MNI}\begin{array}{l}
D(f_{(X_1,X_2)|x_3},f_{(X_1,X_2)})=\diy \Hw_{\phi_{(X_1,X_2)|x_3}}(X_1,X_2)-H((X_1,X_2)|x_3)\\
\quad= \diy \frac{1}{2}\log \left((2\pi)^2 |\Sigma_1|\right)-\frac{1}{2}\log \left((2 \pi)^2 |\overline{\Sigma}|\right)\\
\qquad+ \diy \frac{1}{2} \sum_{i,j=1,2}\Sigma_{ij}^{*^{-1}} \left\{ {\overline{\Sigma}}_{ij}
+\overline{\mu}_i \overline{\mu}_j-\mu_j \overline{\mu}_i-\mu_i \overline{\mu}_j+ \mu_i\mu_j\right\}.
\end{array}\end{eqnarray}
Note that here $\phi_{(X_1,X_2)|x_3}(x_1,x_2)=f(x_1,x_2|x_3)\Big/f(x_1,x_2)$. For simplification and avoiding confusion we introduce $\Sigma_{ij}^{*^{-1}}$
 as the cells in the concentration matrix $\Sigma_1^{-1}$.\\

The conditional DE $H((X_1,X_2)|x_3)$ follows directly from (\ref{multivariate entropy}) and
\beqq (X_1,X_2|X_3=x_3)\sim \mathcal{N}(\overline{\mu},\overline{\Sigma}).\eeqq
On the other hand, for $\Hw_{\phi_{(X_1,X_2)|x_3}}(X_1,X_2)$ we have the respective formula:
\beqq\begin{array}{l}
\Hw_{\phi_{(X_1,X_2)|x_3}}(X_1,X_2)\\
\quad=\diy\log\left((2\pi)|\Sigma_1|^{\frac{1}{2}}\right)+\frac{1}{2} \sum_{i,j=1,2}\Sigma_{ij}^{*^{-1}}
\int_{\bbR^2} f(x_1,x_2|x_3) (x_i-\mu_i)(x_j-\mu_j) \rd x_1\rd x_2\\
\quad=\diy \log\left((2\pi)|\Sigma_1|^{\frac{1}{2}}\right)
+\frac{1}{2} \sum_{i,j=1,2}\Sigma_{ij}^{*^{-1}}\bigg\{E\left[X_iX_j|X_3\right]-
\mu_j E\left[X_i|X_3\right]-\mu_i E\left[X_j|X_3\right]+\mu_i\mu_j\bigg\}\\
\quad=\diy \log\left((2\pi)|\Sigma_1|^{\frac{1}{2}}\right)+\frac{1}{2} \sum_{i,j=1,2}\Sigma_{ij}^{*^{-1}}\Big\{\overline{\Sigma}_{ij}
+\overline{\mu}_i \overline{\mu}_j-\mu_j\overline{\mu}_i
-\mu_i \overline{\mu}_j+\mu_i\mu_j\Big\}.\end{array}\eeqq

In addition, according to (\ref{conditionalWK-B}), we are entitled to give a comprehensive expression\\
for $D^{\rm w}_\phi(f_{(X_1,X_2)|x_3},f_{(X_1,X_2)})$. Define
\beq\label{Theta.Lambda}\begin{array}{l}
\Theta(x_3)=\diy E\left[\prod\limits_{i=1}^2 (X_i-\mu_i)^2|X_3\right]\\
\overline{\Lambda}_{ij}=\diy E\bigg[\prod\limits_{k=1}^2(X_k-\mu_k)^2(X_i-\omu_i)(X_j-\omu_j)|X_3\bigg].\end{array}\eeq
Then we get
\beq\label{EWB} \Hw_{\phi}(X_1,X_2|X_3=x_3)=\diy\left[ \frac{1}{2} \log \left((2\pi)^2 |\overline{\Sigma}|\right)\right]\;\Theta(x_3)
+\frac{1}{2}\sum_{i=1}^2 \sum_{j=1}^2 \overline{\Sigma}_{ij}^{-1}\overline{\Lambda}_{ij}.\eeq
We confine in the Appendix the calculations related to the precise value of $\Theta(x_3)$ and $\overline{\Lambda}_{ij}$.\\
Now for WF
\beqq \phi'_{(X_1,X_2)|x_3}(x_1,x_2)=\diy\prod\limits_{i=1}^2 (x_i-\mu_i)^2\bigg[\frac{f(x_1,x_2|x_3)}{f(x_1,x_2)}\bigg],\eeqq
we draw the reader's attention to the following assertion:
\beq \label{EWCB}\begin{array}{ccl}
\Hw_{\phi'_{(X_1,X_2)|x_3}}(X_1,X_2)=\diy \frac{1}{2}\log \left[(2\pi)^2|\Sigma_1|\right] \Theta(x_3)+
\diy\frac{1}{2}\sum_{i=1}^2\sum_{j=1}^2 \Sigma_{ij}^{*^{-1}}\Upsilon_{ij}.\end{array}\eeq
Here $\Theta(x_3)$ is defined as in (\ref{Theta.Lambda}) and
\beq \Upsilon_{ij}=\diy E\bigg[\prod\limits_{k=1}^2(X_k-\mu_k)^2 (X_i-\mu_i)(X_j-\mu_j)|X_3\bigg].\eeq
As before the explicit expressions of $\Theta(x_3)$ and $\Upsilon_{ij}$ are given in Appendix.

Following Corollary \ref{cor.1}, combine (\ref{EWB}) and (\ref{EWCB}), and obtain the following quite long expression for the mutual WE:
\beq \label{IWMN}\begin{array}{l}
\diy D^{\rm w}_\phi(f_{(X_1,X_2)|x_3},f_{(X_1,X_2)})\\
\hspace{0.1cm}
\quad=\diy \frac{1}{2}\log \left[|\Sigma_1|\big/|\OSigma|\right] \Theta(x_3)
+\diy\frac{1}{2}\sum_{i=1}^2\sum_{j=1}^2 \Sigma_{ij}^{*^{-1}}\Upsilon_{ij}
-\frac{1}{2}\sum_{i=1}^2 \sum_{j=1}^2 \overline{\Sigma}_{ij}^{-1}\overline{\Lambda}_{ij}.
\end{array}\eeq
Consequently, it is now clear, both Kullbak-Leibler and weighted Kulback-Leibler information measures
for Gaussian conditional pair $(X_1,X_2)|X_3=x_3)$ precisely depend on mean and
covariances between all random variables individually. Because of this fact, it is logical
to wonder about the effect of correlations on them and whether this effect on kullback-Leibler information
is completely analogous to the weighted one.

One obviously can imagine if the dependency between $(X_1,X_2)$
and $X_3$ is increasing then the density function of conditional vector $(X_1,X_2)|X_3=x_3)$ becomes more
far than join density function $(X_1,X_2)$, on the other way we understand that knowing dependent random
variable $X_3$ gives more information.\\
Indeed to prove this claim we shall present more evidences,
so that two particular examples are considering in the following.
\begin{exam}\label{exam1}{\rm
In Gaussian case assume $\underline{\mu}=\underline{0}$ and $\underline{\Sigma}=\left[
                                                                                \begin{array}{ccc}
                                                                                  1 \ &\  \rho\  & \ \rho^2 \\
                                                                                  \rho \ &\  1 \ &\  0 \\
                                                                                  \rho^2\  &\  0\  &\  1 \\
                                                                                \end{array}
                                                                              \right]$, then
\beqq (X_1,X_2|X_3=x_3)\sim \mathcal{N}(\overline{\mu},\overline{\Sigma}),\eeqq
where
\beqq\overline{\mu}=\left[
                   \begin{array}{c}
                     \rho^2 x_3\\
                     0 \\
                   \end{array}
                 \right] \ \ \ \ \hbox{and}\ \ \ \ \overline{\Sigma}=\left[
                                                              \begin{array}{cc}
                                                                1-\rho^4 & \rho \\
                                                                \rho & 1 \\
                                                              \end{array}
                                                            \right].\eeqq
Following $\Sigma_1$ as before, i.e. the covariance matrix for pair $(X_1,X_2)$, one yields $|\Sigma_1|=1-\rho^2$, $|\overline{\Sigma}|=1-\rho^2-\rho^4$ and
\begin{eqnarray*}\begin{array}{l}
\diy\sum_{i,j=1,2}\Sigma_{ij}^{*^{-1}}\left\{\overline{\Sigma}_{ij}+\overline{\mu}_i \overline{\mu}_j-\mu_j\overline{\mu}_i
-\mu_i \overline{\mu}_j+\mu_i\mu_j\right\}=\diy \sum_{i,j=1,2}\Sigma_{ij}^{*^{-1}}\left\{\overline{\Sigma}_{ij}
+\overline{\mu}_i \overline{\mu}_j\right\}\\
\quad=\diy\Sigma_{11}^{*^{-1}}\left\{ \overline{\Sigma}_{11}+\overline{\mu}_1^2\right\}+\Sigma_{12}^{*^{-1}}\overline{\Sigma}_{12}
+\Sigma_{21}^{*^{-1}}\overline{\Sigma}_{21}+
\Sigma_{22}^{*^{-1}}\overline{\Sigma}_{22}=\diy\frac{2-2\rho^2+\rho^4(x_3^2-1)}{1-\rho^2}.\end{array}
\end{eqnarray*}
Using the above expression in (\ref{MNI}), we obtain
\begin{eqnarray}\label{eqqqq1}
D(f_{(X_1,X_2)|x_3},f_{(X_1,X_2)})=\frac{1}{2} \log \left[\frac{1-\rho^2}{(1-\rho^2-\rho^4)}\right]+\frac{\rho^4}{2(1-\rho^2)}.(x_3^2-1)+1.
\end{eqnarray}

Taking look into this equation, we are not explicitly able to realize if it is a monotonic function with respect to $\rho$,
but obviously it is an even function with respect to $\rho$ and $x_3$, that is the relative DE does not depend on
the sign of correlation coefficient $\rho$. Although from (\ref{eqqqq1}) it is clear that the absolute value of $x_3$ effects on
the information.
\begin{center}
{\includegraphics[scale=.8]{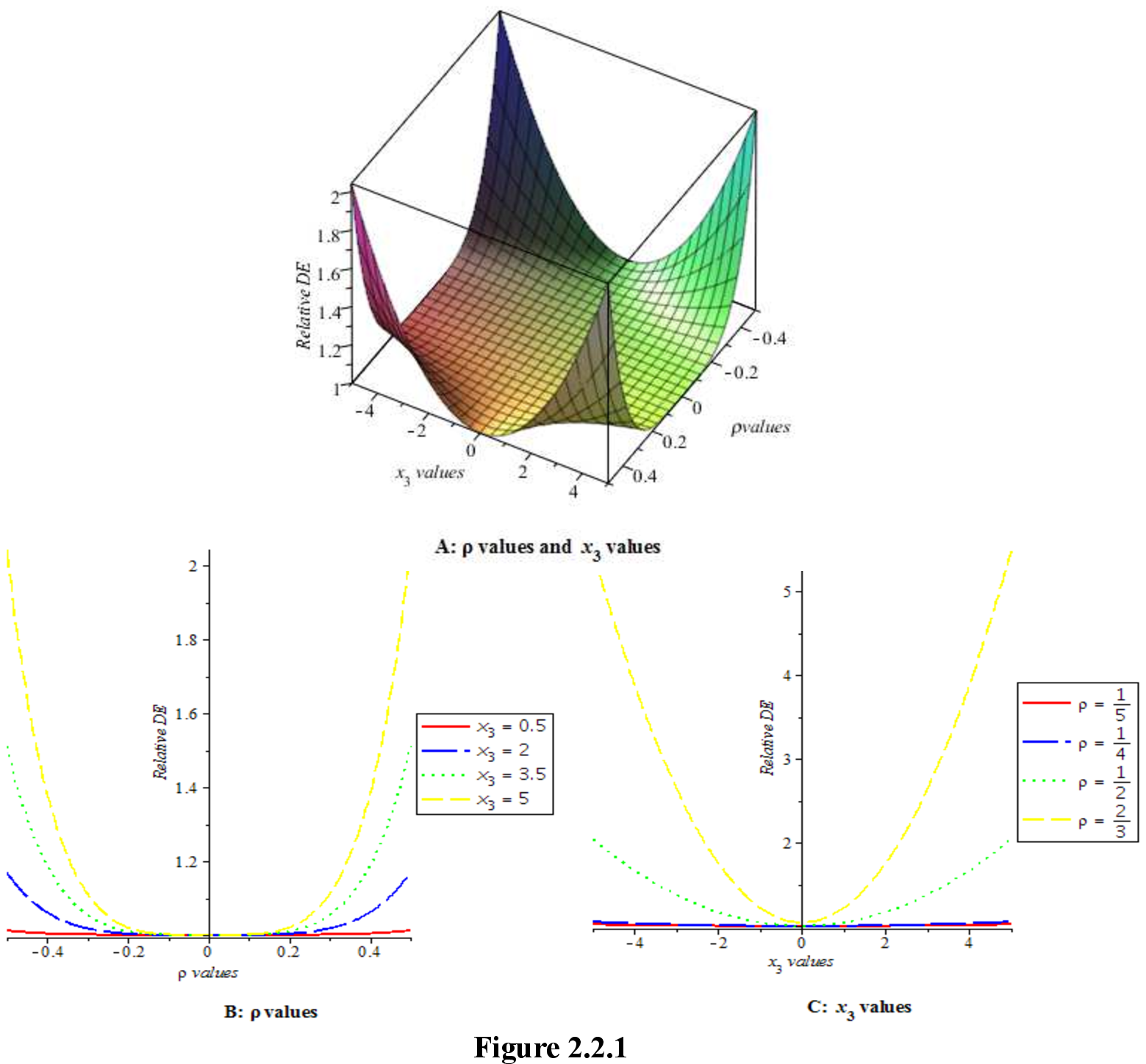}}
\end{center}
\label{fig}
In Figure 2.2.1 we see that first the relative DE takes non-negative values (Gibbs inequality, see \cite{cover2006, KS}). Second the information raises when absolute value of $\rho$ and $x_3$ is increasing which is completely coincide with our expected claim, observe Figures 2.2.1(B) and 2.2.1(C).  On the other hand in this example the dependency between $X_1$ and $X_3$, $\rho^2$, and the discrimination between $f_{(X_1,X_2)|x_3}$, $f_{(X_1,X_2)}$ change in the same direction. Although since we have concentrated on the information between distributions $(X_1,X_2)|X_3=x_3)$ and $(X_1,X_2)$, the correlation between $X_1$ and $X_2$ is not in our attention.\\

Now, we shall switch to the relative WE in order to discover if we can extend the similar impression for the weighted one.
It is worthwhile nothing that since in weighted information apart of probabilities we also add the amount values of
RVs $X_1$, $X_2$, thus probably our perception changes.\\

Consider $\mu$ and $\Sigma$ as already given in Example \ref{exam1}, we get
\beq\begin{array}{cl}
\diy \Theta(x_3)=\diy 1+2\rho^2+\rho^4(x_3^2-1),\\
\diy \overline{\Lambda}_{ij}=\diy \alpha_{ij}(\rho)+\rho^4 x_3^2\bigg((1-\rho^4)\OSigma_{ij}+2\OSigma_{1i}\OSigma_{1j}\bigg).
\end{array}\eeq
here
\beqq\diy \alpha_{ij}(\alpha)=\diy (1-\rho^4)\bigg(\OSigma_{ij}+2\OSigma_{2i}\OSigma_{2j}\bigg)+2\rho\bigg(\rho\OSigma_{ij}+
\OSigma_{1i}\OSigma_{2j}+\OSigma_{1j}\OSigma_{2i}\bigg)\\
\qquad+\diy \OSigma_{1i}\bigg(2\rho\OSigma_{2j}+\OSigma_{1j}\bigg)+\OSigma_{1j}\bigg(2\rho \OSigma_{2i}+\OSigma_{1i}\bigg).\eeqq
Set $\beta_{ij}(\rho)=(\omu_i-\mu_i)(\omu_j-\mu_j)$, we have
\beq\begin{array}{l} \Upsilon_{ij}=\alpha_{ij}(\rho)+\beta_{ij}(\rho).\bigg[1+2\rho^2+\rho^4(x_3^2-1)\bigg]+2\rho^2 x_3(\omu_j-\mu_j).\bigg(\OSigma_{1i}+2\rho\OSigma_{2i}\bigg)\\
\qquad+2\rho^2 x_3(\omu_i-\mu_i).\bigg(\OSigma_{ij}+2\rho\OSigma_{2j}\bigg)+(\rho^4x_3^2).\bigg(\OSigma_{ij}+2\OSigma_{2i}\OSigma_{2j}\bigg).
\end{array}\eeq
By virtue of (\ref{IWMN}), we obtain
\beq\begin{array}{l}\diy D^{\rm w}_\phi(f_{(X_1,X_2)|x_3},f_{(X_1,X_2)})\\
\quad=\frac{1}{2}\log\bigg[\diy\frac{1-\rho^2}{1-\rho^2-\rho^4}\bigg].\bigg(\diy 1+2\rho^2+\rho^4(x_3^2-1)\bigg)\\
\qquad+\diy\frac{1}{2\;(1-\rho^2)}\bigg(3(1-\rho^4)^2+3(1-\rho^4)+6\rho^2-6\rho^4-6\rho^6x_3^2+9\rho^4x_3^2-6\rho^8x_3^2+\rho^8x_3^2\bigg)\\
\qquad-\diy \frac{1}{2\;(1-\rho^2-\rho^4)}\bigg(6\rho^2(1-\rho^4)+6(1-\rho^4)^2+4\rho^4(1-\rho^4)x_3^2-12\rho^4-4\rho^6(1-\rho^4)x_3^2\bigg).
\end{array}\eeq

The following theorem was presented in \cite{suhov2014}:

\begin{thm}\label{RWE Gibbs} {\rm{(The weihghted Gibbs inequality)}}
Given non-negative functions $f$, $g$ and $\phi$, assume the bound
\beq\label{cond.Gibbs}\diy\int\limits_{\bbR^n}\phi (\ux )\big[f(\ux)-g(\ux)\big]\rd \ux\geq 0.\eeq
Then
 $$D^{\rm w}_\phi(f\|g)\geq 0,$$
with equality iff $g\equiv f$.
\end{thm}

The condition (\ref{cond.Gibbs}) is re-written as
\beqq \begin{array}{l}\diy\int\limits_{\bbR^2}\prod\limits_{i=1}^2(x_i-\mu_i)^2\bigg[f(x_1,x_2|x_3)-f(x_1,x_2)\bigg]\rd x_1\rd x_2\\
\quad=\diy E\bigg[\prod\limits_{i=1}^2(X_i-\mu_i)^2|X_3\bigg]-E\bigg[\prod\limits_{i=1}^2(X_i-\mu_i)^2\bigg]\geq 0.\end{array}\eeqq
Correspondingly in this example we obtain:
$$\Theta(x_3)-E\bigg[\prod\limits_{i=1}^2 X_i^2\bigg]=\rho^4 (x_3^2-1).$$
This states that for $x_3\in(-1,1)$ the condition (\ref{cond.Gibbs}) is violated and there is no guaranty that the relative WE takes non-negative values whereas $D^{\rm w}_\phi(f_{(X_1,X_2)|x_3},f_{(X_1,X_2)})\geq 0$, see Figure 2.2.2(A). Further, a similar pattern as the relative DE for the relative WE's behavior is confirmed on Figure 2.2.2(B) and 2.2.2(C) with the same values of $\rho$ and $x_3$.
\begin{center}
{\includegraphics[scale=.8]{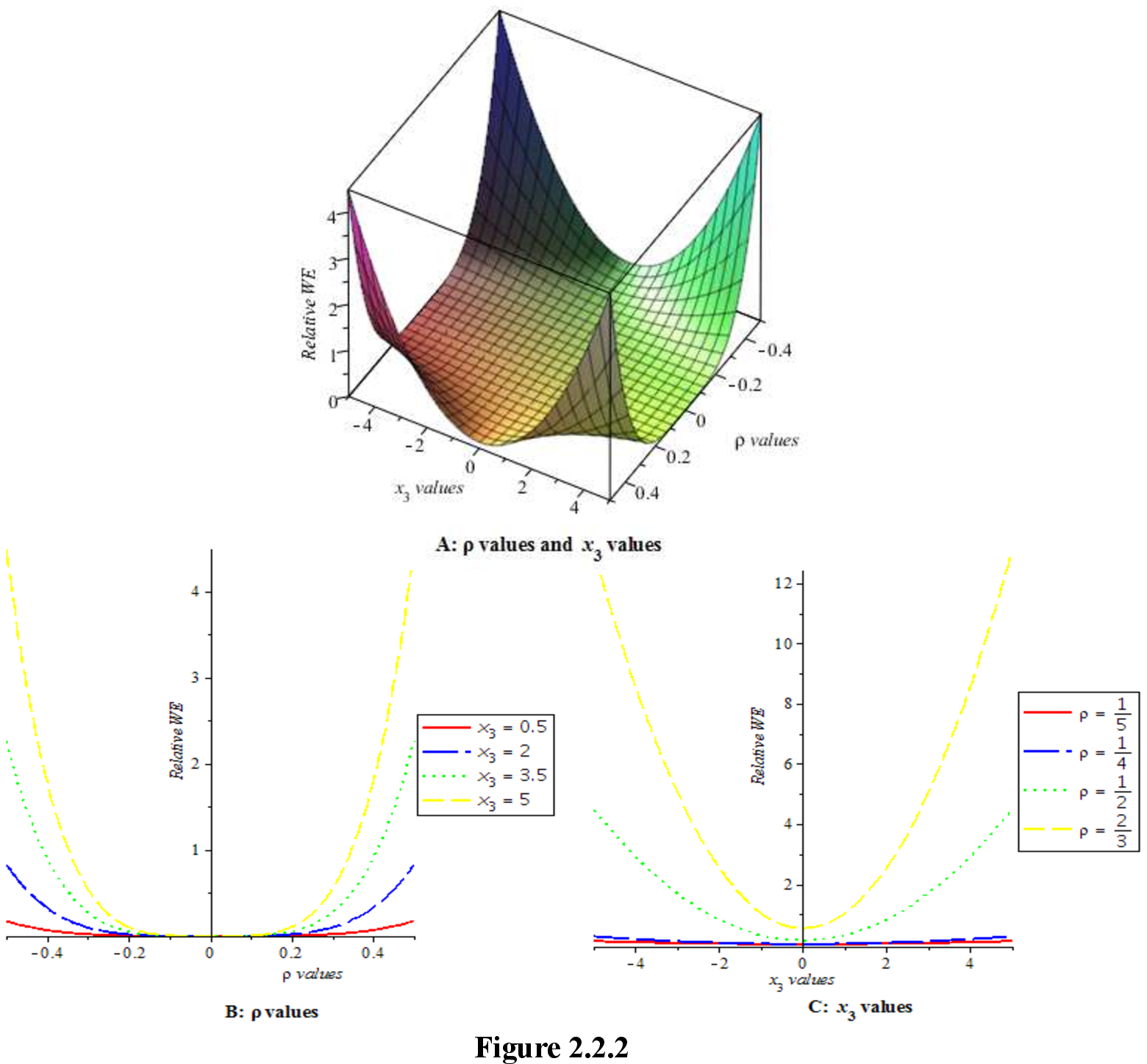}}
\end{center}
\label{fig}

}\end{exam}
Our observations still are preliminary, and we think that further examples are needed here, to build a detailed picture.
Hence let us now devote our efforts on another special case of Gaussian distribution which has been called from Example 3.4.1 page 39, \cite{T}:
\vskip 1 truecm
 \begin{exam}\label{exam2}{\rm
Let $\underline{X}=(X_1,X_2,X_3)$ be distributed according to an $\mathcal{N}(\underline{0}, \Sigma)$ distribution,
where $\Sigma_{ii}=1, (i=1,2,3)$ and $\Sigma_{12}=1-2\rho$, $\Sigma_{13}=\Sigma_{23}=1-\rho$, $0<\rho<\frac{1}{2}$.
For every fix $\underline{C}=(C_1,C_2, C_3)^T\in \mathbb{R}^3$. we can write,
\begin{eqnarray*}
C^T\Sigma C= (1-\rho)(C_1+C_2+C_3)^2+\rho(C_1+C_2)^2+\rho C_3^2,
\end{eqnarray*}
Since $C^T \Sigma C \geq 0$ holds, and the equality holds if and only if $C_1=C_2=C_3=0$, so $\Sigma$ is a positive define matrix.
We have then,
$$\overline{\mu}=\left[
                   \begin{array}{c}
                     (1-\rho) x_3\\
                     (1-\rho) x_3 \\
                   \end{array}
                 \right] \ \ \ \ ,\ \ \ \ \overline{\Sigma}=\left[
                                                              \begin{array}{cc}
                                                                \rho(2-\rho) & -\rho^2 \\
                                                                -\rho^2 & \rho(2-\rho) \\
                                                              \end{array}
                                                            \right].$$
Owing to (\ref{MNI}) we first calculate,
\begin{eqnarray*}\begin{array}{l}
\diy \sum_{i,j=1,2}\Sigma_{ij}^{*^{-1}} \left\{ {\overline{\Sigma}}_{ij}+\overline{\mu}_i \overline{\mu}_j-\mu_j \overline{\mu}_i-\mu_i \overline{\mu}_j+
 \mu_i\mu_j\right\}\\
 \quad =\diy \sum_{i,j=1,2}\Sigma_{ij}^{*^{-1}} \left\{ {\overline{\Sigma}}_{ij}+\overline{\mu}_i \overline{\mu}_j\right\}
=1+\rho+(1-\rho)x_3^2.\end{array}
\end{eqnarray*}
\begin{center}
{\includegraphics[scale=.8]{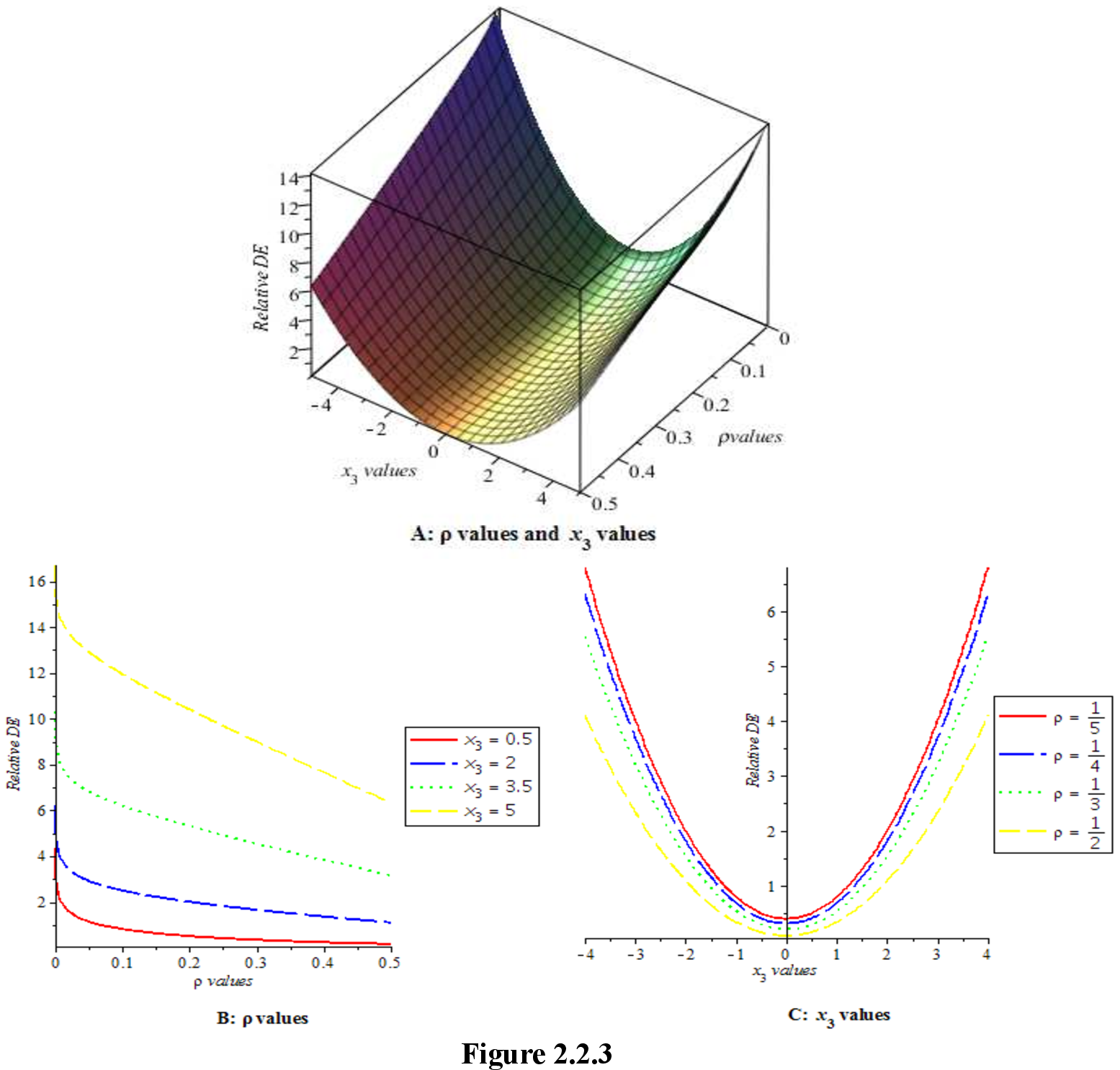}}
\end{center}
\label{fig}
Therefore, for $\rho\in(0,\frac{1}{2})$, we drive
\begin{eqnarray}\begin{array}{l}
D(f_{(X_1,X_2)|x_3},f_{(X_1,X_2)})\\
\quad=\diy\frac{1}{2} \log \left[(2\pi)^2 4\rho (1-\rho)\right]-\frac{1}{2}
\log \left[(2\pi)^2 4\rho^2(1-\rho)\right]+\frac{1}{2}\left[1+\rho+(1-\rho)x_3^2\right]\\
\quad=\diy \frac{1}{2}\left[1+\rho+(1-\rho)x_3^2-\log(\rho)\right]-1.\end{array}
\end{eqnarray}

Next, Figure 2.2.3 shows a more interesting character of behavior. Function $D(f_{(X_1,X_2)|x_3},f_{(X_1,X_2)})$ takes non-negative values. However the relative DE decreases in $\rho\in(0,\diy\frac{1}{2})$. In other word, this is another example showing our perception holds true: when the dependency within RVs $X_1$ and $X_3$ raises the information increases as well. \\

Furthermore, the following expression for $\Theta(x_3)$ emerges:
\beq\label{theta.exam2}
\Theta(x_3)=\rho^2(2-\rho)^2+4\rho^4+2\rho(2-\rho)(1-\rho)^2 x_3^2-4\rho^2(1-\rho)^2x_3^2+(1-\rho)^4 x_3^4.
\eeq
Therefore owing to (\ref{IWMN}), one yields
\beq \begin{array}{l}
\diy D^{\rm w}_\phi(f_{(X_1,X_2)|x_3},f_{(X_1,X_2)})=\\
\quad \diy\frac{1}{2}\log\bigg[(2\pi)^2\bigg(\frac{4\rho(1-\rho)}{\rho^2(2-\rho)^2-\rho^4}\bigg)\bigg].\Theta(x_3)\\
\qquad\diy+\frac{1}{8\rho(1-\rho)}\bigg((\Upsilon_{11}+\Upsilon_{22})+(2\rho-1)(\Upsilon_{12})\bigg)\\
\qquad\diy-\frac{1}{2(\rho^2(2-\rho)^2-\rho^4)}\bigg(\rho(2-\rho)(\overline{\Lambda}_{11}+\overline{\Lambda}_{22})
+2\rho^2(\overline{\Lambda}_{12})\bigg).\end{array}\eeq
Here $\Theta(x_3)$ is as (\ref{theta.exam2}) and we have written the open form of $\overline{\Lambda}_{ij}$ and $\Upsilon_{ij}$ in Appendix.

Let us now check the statues of the condition (\ref{cond.Gibbs}):
\beq\label{cond11}\begin{array}{l}\diy \Theta(x_3)-\bigg(1+(1-2\rho)^2\bigg)\\
\qquad=\diy \rho^2(2-\rho)^2+4\rho^4+2\rho(2-\rho)(1-\rho)^2x_3^2-4\rho^2(1-\rho)^2x_3^2+(1-\rho)^4x_3^4-1-(1-2\rho)^2.\end{array}\eeq
Analyzing (\ref{cond11}), one can explore that the condition (\ref{cond.Gibbs}) doesn't hold true for all values of $\rho$ and $x_3$ whereas as Figure 2.2.4 shows, the relative WE is non-negative (within the indicated range of $(\rho,x_3)$.

Finally, the plots given in Figure 2.2.4(B) and 2.2.4(C) give an impression that the behavior of  $ D^{\rm w}_\phi(f_{(X_1,X_2)|x_3},f_{(X_1,X_2)})$ is more complicated. Other words, in this example the information doesn't behave monotonically with respect to $\rho$ and $x_3$. Consequently, in spite of standard case, in weighted form one does not yield that the dependency between $X_1$ and $X_3$ effects directly on the information.
\begin{center}
{\includegraphics[scale=.8]{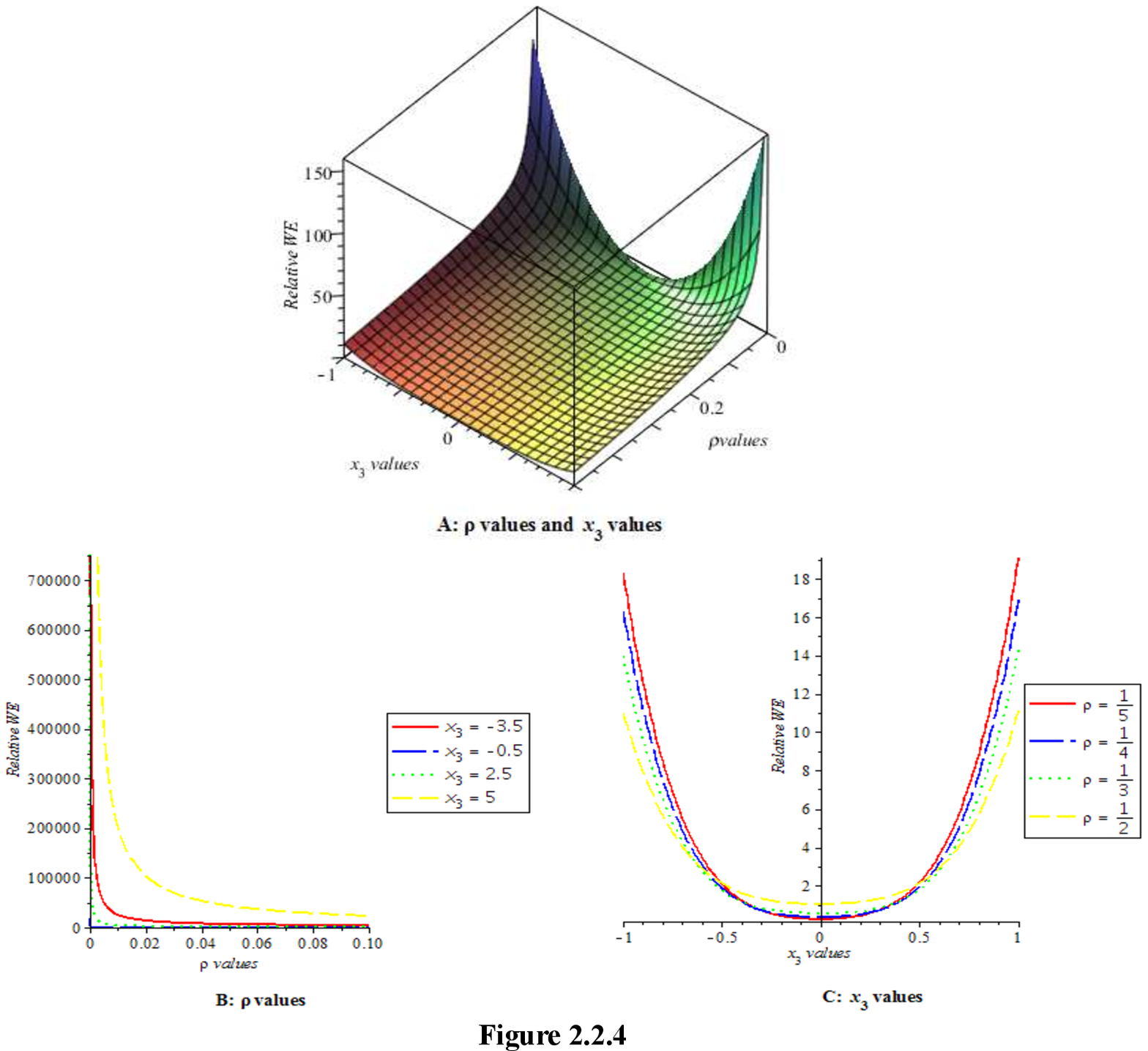}}
\end{center}
\label{fig}

}\end{exam}

\section{An application: Weighted deviance information criterion}
\def\ty{\tilde{y}}
Concluding this paper in this section, we briefly demonstrate an application of the relative DE and WE by exploiting Beysian analysis in model selecting, cf. \cite{zelner1971, yasaei2014}.

Assume that $f(\ty)$ and $g(\ty)$ respectively represent the PDFs of the "true model" and the "approximating model" on the same measurable space. For given WF $\phi$, the relative WE or weighted Kullback-Leibler divergence is given by:
\beq\label{Eq:3.1} D^{\rm w}_\phi(f\|g)=\bbE_{\ty}\big[\phi(\ty)\log f(\ty)\big]-\bbE_{\ty}\big[\phi(\ty)\log g(\ty)\big].\eeq
Note that such a quantity is not always non-negative. Namely the smaller the value of $D^{\rm w}_\phi$, the closer we consider the model $g$ to be the true distribution. Hence in practice the first part of (\ref{Eq:3.1}) is negligible in model comparison for given data $\by=(y_1,\dots,y_n)$ with weighs $\phi(\by)$.\\
As $n$ increases to infinity, the following expression, weighted log-likelihood (say):
\beqq \frac{1}{n}L^{\rm w}_\phi(\theta|y):=\frac{1}{n} \sum\limits_{i=1}^n \log \Big[g(y_i|\theta)^{\phi(y_i)}\Big]\eeqq
tends to $\bbE_{\ty}\big[\phi(\ty)\log g(\ty|\theta)\big]$ by the law of large numbers. Here $\phi(y_i)$ is the weight for $y_i$ and $\ty$ is supposed to be an unknown but potentially observable quantity coming from the same distribution $f$ and independent of $y$.\\

Next in agreement with \cite{S.et.al}, we propose the weighted deviance information criterion (WDIC):
\beq\label{WDIC} WDIC= DE^{\rm w}_\phi(\hat{\theta},y)+2 p^{\rm w}_D,\eeq
as an adaptation of the Akaike information criterion for weighted case for Bayesian models. Consider the penalty of over-estimating $p^{\rm w}_D$ by
\beq \label{penalty} p^{\rm w}_D=\bbE_{\theta|y}\Big[DE^{\rm w}_\phi(\theta,y)\Big]-DE^{\rm w}_\phi(\hat{\theta},y)\eeq
in order to estimate the "effective number of parameters". Here
$$DE^{\rm w}_\phi(\theta,y)=-2 \sum \limits_{i=1}^n \log \Big(g(y_i|\theta)^{\phi(y_i)}\Big).$$
As far the full model specification of Bayesian statistics contains a prior function $\Pi(\theta)$ in addition to the likelihood, and the inference can be derived from the posterior distribution $\Pi(\theta|y)\propto L(\theta|y)\Pi(\theta)$, therefore $\hat{\theta}$ could be either posterior mean or mode. In practice the advantage of WDIC with respect to DIC is observed when the data has the utility (weight) non equal to one.
\vskip .5 truecm
{\bf Remark:} It would be interesting to investigate some simulation results as evidence of this fact by using Markov chain Monte Carlo (MCMC) method. This also is one of our intentions for future works.
\section{APPENDIX}
{\bf Proof of (\ref{weightednormal}):}\\

According to (\ref{eq:WDE0}) for the Gaussian PDF and given WF $\phi(\ux)=\prod\limits_{i=1}^3 (x_i-\mu_i)^2$, one can write:
\begin{eqnarray}\label{AP1}\begin{array}{l}
\Hw_\phi(\UX)=
\diy\left[ \frac{1}{2} \log \left((2\pi)^3 |\Sigma|\right)\right]E\left[\prod_{k=1}^3 (X_k-\mu_k)^2\right]\\
\qquad\qquad+\diy \frac{1}{2}\diy \int_{\mathbb{R}^{3}}\prod_{k=1}^3 (x_k-\mu_k)^2 f(\underline{x}) \sum_{i=1}^3 \sum_{j=1}^3 (x_i-\mu_i)\Sigma_{ij}^{-1}(x_j-\mu_j)
\rd\underline{x}\\
\qquad\quad = \diy\left[ \frac{1}{2} \log \left((2\pi)^3 |\Sigma|\right)\right]E\left[\prod_{k=1}^3 (X_k-\mu_k)^2\right]\\
\qquad\qquad+\diy\frac{1}{2}\sum_{i=1}^3 \sum_{j=1}^3 \Sigma_{ij}^{-1}
E\left[ \prod_{k=1}^3 (X_k-\mu_k)^2 (X_i-\mu_i)(X_j-\mu_j)\right].\end{array}
\end{eqnarray}

Set $y_i=x_i-\mu_i$, then $y_i \sim \mathcal{N}(0, \Sigma_{ij})$ and moreover for odd $M=r_1+r_2+\dots+r_n$, $E[\prod_{i=1}^n y_i^{r_i}]=0$. Now let us focus on the last expectation in (\ref{AP1}) which takes the form:
\beqq \begin{array}{l}
\Lambda_{ij}:=E\left[ \prod_{k=1}^3 (X_k-\mu_k)^2(X_i-\mu_i)(X_j-\mu_j)\right]=\diy E\left[ \prod_{k=1}^3 Y_k^2 Y_i Y_j\right]\\
\quad=\diy E\big[Y_1^2Y_2^2\big]E\big[Y_3^2\big]E\big[Y_iY_j\big]+E\big[Y_1^2Y_2^2\big]E\big[Y_3Y_i\big]E\big[Y_3Y_j\big]+E\big[Y_1^2Y_2^2\big]
E\big[Y_3Y_j\big]E\big[Y_3Y_i\big]\\
\quad+\diy E\big[Y_1^2Y_3\big]E\big[Y_2^2Y_3\big]E\big[Y_iY_j\big]+E\big[Y_1^2Y_3\big]E\big[Y_2^2Y_i\big]E\big[Y_3Y_j\big]+E\big[Y_1^2Y_3\big]
E\big[Y_2^2Y_j\big]E\big[Y_3Y_i\big]\\
\quad+\diy E\big[Y_1^2Y_3\big]E\big[Y_2^2Y_3\big]E\big[Y_iY_j\big]+E\big[Y_1^2Y_3\big]E\big[Y_2^2Y_i\big]E\big[Y_3Y_j\big]+E\big[Y_1^2Y_3\big]
E\big[Y_2^2Y_j\big]E\big[Y_3Y_i\big]\\
\quad+E\big[Y_1^2Y_i\big]E\big[Y_2^2Y_3\big]E\big[Y_3Y_j\big]+E\big[Y_1^2Y_i\big]E\big[Y_2^2Y_3\big]E\big[Y_3Y_j\big]
+E\big[Y_1^2Y_i\big]E\big[Y_2^2Y_j\big]E\big[Y_3^2\big]\\
\quad+E\big[Y_1^2Y_j\big]E\big[Y_2^2Y_3\big]E\big[Y_3Y_i\big]+E\big[Y_1^2Y_j\big]E\big[Y_2^2Y_3\big]E\big[Y_3Y_i\big]+E\big[Y_1^2Y_j\big]
E\big[Y_2^2Y_i\big]E\big[Y_3^2]\\
\\
\quad=\diy \bigg(\Sigma_{11}\Sigma_{22}+2\big(\Sigma_{12}\big)^2\bigg).\bigg(\Sigma_{33}\Sigma_{ij}+2\;\Sigma_{3i}\Sigma_{3j}\bigg).
\end{array}\eeqq
Next, in (\ref{AP1}) we need to find one more expectation:
\beq \begin{array}{l} \Xi:= E\left[\prod_{k=1}^3 (X_k-\mu_k)^2\right]=E\left[\prod_{k=1}^3 Y_k^2\right]=
E\big[Y_1^2\big].\bigg( E\big[Y_2^2\big]E\big[Y_3^2\big]
+2\;\big(E\big[Y_2Y_3\big]\big)^2\bigg)\\
\quad+2\;E\big[Y_1Y_2].\bigg(E\big[Y_1Y_2\big]E\big[Y_3^2\big]+2\;E\big[Y_1Y_3\big]E\big[Y_2Y_3\big]\bigg)\\
\quad+2\;E\big[Y_1Y_3\big]\bigg(2\;E\big[Y_1Y_2\big]E\big[Y_2Y_3\big]+E\big[Y_1Y_3\big]E\big[Y_2^2\big]\bigg)\\
\quad=\diy \Sigma_{11}\bigg(\Sigma_{22}\Sigma_{33}+2\;(\Sigma_{23})^2\bigg)+2\;\Sigma_{12}\bigg(\Sigma_{12}\Sigma_{33}+2\Sigma_{13}\Sigma_{23}\bigg)\\
\qquad+2\;\Sigma_{13}\bigg(2\Sigma_{12}\Sigma_{23}+\Sigma_{13}\Sigma_{22}\bigg).
\end{array}\eeq

By replacing the above expressions in (\ref{AP1}) we can obtain
\beq \Hw_{\phi}(\UX)=\diy\left[ \frac{1}{2} \log \left((2\pi)^3 |\Sigma|\right)\right]\;\Xi
+\frac{1}{2}\sum_{i=1}^3 \sum_{j=1}^3 \Sigma_{ij}^{-1}\Lambda_{ij}.\eeq
which is exactly what we are looking for. $\quad$ $\Box$

\vskip .5 truecm

{\bf Proof of (\ref{EWB})}:\\

Recall the conditional WE:
\beqq\begin{array}{l}
\Hw_{\phi}(X_1,X_2|X_3=x_3)\\
\quad=-\diy\int \phi(x_1,x_2)f(x_1,x_2|x_3)\log f(x_1,x_2|x_3) \rd x_1\;\rd x_2\\
\quad=\diy \frac{1}{2}\log \left[(2\pi)^2|\OSigma_1|\right] E\left[\prod\limits_{i=1}^2 (X_i-\mu_i)^2|X_3\right]\\
\qquad+\diy\frac{1}{2}\sum_{i,j=1,2} \OSigma_{ij}^{-1}\int\prod\limits_{k=1}^2(x_k-\mu_k)^2(x_i-\omu_i)(x_j-\omu_j) f(x_1,x_2|x_3)\rd x_1\;\rd x_2.\end{array}\eeqq
Observe that
\beq \begin{array}{l} E\left[\prod\limits_{i=1}^2 (X_i-\mu_i)^2|X_3\right]\\
\qquad\quad=\diy E\bigg[\prod\limits_{i=1}^2 \Big((X_i-\overline{\mu}_i)^2+2(X_i-\overline{\mu}_i)
(\overline{\mu}_i-\mu_i)+(\overline{\mu}_i-\mu_i)^2\Big)\big|X_3\bigg]\\
\qquad\quad=\diy E\big[\OY_1^2\OY_2^2\big]+E\big[\OY_1^2\big](\omu_2-\mu_2)^2+ E\big[\OY_2^2\big](\omu_1-\mu_1)^2\\
\qquad\qquad+\diy 4\;E\big[\OY_1\OY_2\big]\prod\limits_{i=1}^2(\omu_i-\mu_i)+\prod\limits_{i=1}^2(\omu_i-\mu_i)^2.\end{array}\eeq
where $(\OY_1,\OY_2)\sim{\rm N}(\underline{0},\overline{\Sigma})$, $\omu$ and $\overline{\Sigma}$ are as in (\ref{Omu}) and (\ref{OSigma}). Therefore
\beq\label{Ap2} \begin{array}{l} E\left[\prod\limits_{i=1}^2 (X_i-\mu_i)^2|X_3\right]=\diy\bigg(\OSigma_{11}\OSigma_{22}+2(\OSigma_{12})^2\bigg)
+\OSigma_{11}\bigg(\Sigma_{23}\Sigma_{33}^{-1}(x_3-\mu_3)\bigg)^2\\
\qquad+\OSigma_{22}\bigg(\Sigma_{13}\Sigma_{33}^{-1}(x_3-\mu_3)\bigg)^2
+\diy 4\;\OSigma_{12}\prod\limits_{i=1}^2 \bigg(\Sigma_{i3}\Sigma_{33}^{-1}(x_3-\mu_3)\bigg)\\
\qquad\quad+\prod\limits_{i=1}^2 \bigg(\Sigma_{i3}\Sigma_{33}^{-1}(x_3-\mu_3)\bigg)^2:=\Theta(x_3) (\hbox{say}).\end{array}\eeq
Furthermore
\beq \label{OLambda}\begin{array}{l}
\overline{\Lambda}_{ij}:=E\bigg[\prod\limits_{k=1}^2(X_k-\mu_k)^2 (X_i-\omu_i)(X_j-\omu_j)|X_3\bigg]\\
\quad=\diy E\big[\OY_1^2\OY_2^2\OY_i\OY_j\big]+\sum\limits_{k=1}^2(\omu_k-\mu_k)^2 E\big[\OY_k^2 \OY_i\OY_j\big]\\
\qquad+\diy \prod\limits_{k=1}^2 (\omu_k-\mu_k)^2E\big[\OY_i\OY_j\big]+4\;\prod\limits_{k=1}^2 (\omu_k-\mu_k)E\big[\OY_1\OY_2\OY_i\OY_j\big].
\end{array}\eeq
\vskip 1 truecm
Owing to $(\OY_1,\OY_2)\sim{\rm N}(\underline{0},\overline{\Sigma})$, we have the following list of assertions:
\beqq \begin{array}{l}
\diy E\big[\OY_1^2\OY_2^2\OY_i\OY_j\big]=\OSigma_{11}\bigg(\OSigma_{22}\OSigma_{ij}+2\OSigma_{2i}\OSigma_{2j}\bigg)
+2\diy\;\OSigma_{12}\bigg(\OSigma_{12}\OSigma_{ij}+\OSigma_{1i}\OSigma_{2j}+\OSigma_{1j}\OSigma_{2i}\bigg)\\
\quad+\diy \OSigma_{1i}\bigg(2\;\OSigma_{12}\OSigma_{2j}+\OSigma_{22}\OSigma_{1j}\bigg)
+\diy \OSigma_{1j}\bigg(2\;\OSigma_{12}\OSigma_{2i}+\OSigma_{22}\OSigma_{1i}\bigg),\end{array}\eeqq
and for $i,j,k=1,2$
\beq\label{Ap3}\begin{array}{cl}\diy E\big[\OY_1\OY_2\OY_i\OY_j\big]=\OSigma_{12}\OSigma_{ij}+\OSigma_{1i}\OSigma_{2j}+\OSigma_{1j}\OSigma_{2i},\;\;\diy E\big[\OY_i^2\OY_j\OY_k\big]=\OSigma_{ii}\OSigma_{jk}+2\;\OSigma_{ij}\OSigma_{ik},\\
\diy E\big[\OY_1^2\OY_2^2\big]=\OSigma_{11}\OSigma_{22}+2(\OSigma_{12})^2,\diy E\big[\OY_i\OY_j\big]=\OSigma_{ij},\;\; \diy E\big[\OY_k^2\big]=\OSigma_{kk}\;\;\hbox{and}\\
\diy \omu_k-\mu_k=\Sigma_{k3}\Sigma_{33}^{-1}(x_3-\mu_3).\end{array}\eeq
which leads directly to the result. $\quad$ $\Box$
\vskip 1 truecm

{\bf Proof of (\ref{EWCB})}:\\

By virtue of the definition of WE with given WF
$$\phi'_{(X_1,X_2)|x_3}=\diy\prod\limits_{i=1}^2 (x_i-\mu_i)^2\bigg[\frac{f(X_1,X_2|x_3)}{f(X_1,X_2)}\bigg],$$
we can write:
\beqq \begin{array}{l}
\diy\Hw_{\phi'_{(X_1,X_2)|x_3}}(X_1,X_2)\\
\qquad=-\diy\int_{\mathbb{R}^{2}}\prod\limits_{i=1}^2 (x_i-\mu_i)^2 f(x_1,x_2|x_3)\log f(x_1,x_2) dx_1 dx_2\\
\qquad=\diy \frac{1}{2}\log \left[(2\pi)^2|\Sigma_1|\right] E\left[\prod\limits_{i=1}^2 (X_i-\mu_i)^2|X_3\right]\\
\qquad\quad+\diy\frac{1}{2}\sum_{i,j=1,2} \Sigma_{ij}^{*^{-1}}E\bigg[\prod\limits_{k=1}^2(X_k-\mu_k)^2 (X_i-\mu_i)(X_j-\mu_j)|X_3\bigg].\end{array}
\eeqq
Here $\Theta(x_3)$ has been calculated already in (\ref{Ap2}). Moreover one yields
\beq \label{Upsilon}\begin{array}{l}
\Upsilon_{ij}:=E\bigg[\prod\limits_{k=1}^2(X_k-\mu_k)^2 (X_i-\mu_i)(X_j-\mu_j)|X_3\bigg]\\
\quad=\diy E\bigg[\prod\limits_{k=1}^2 \big(\OY_k^2+2\OY_k(\omu_k-\mu_k)+(\omu_k-\mu_k)^2\big).\big(\OY_i+(\omu_i-\mu_i)\big)
\big(\OY_j+(\omu_j-\mu_j)\big)\bigg].\end{array}\eeq
Applying RV $(\OY_1,\OY_2)$, (\ref{Upsilon}) becomes
\beq \begin{array}{l}\Upsilon_{ij}=\diy E\big[\OY_1^2\OY_2^2\OY_i\OY_j\big]+E\big[\OY_1^2\OY_2^2\big](\omu_i-\mu_i)(\omu_j-\mu_j)\\
\quad+2\diy\;E\big[\OY_1^2\OY_2\OY_i\big](\omu_2-\mu_2)(\omu_j-\mu_j)+2\;E\big[\OY_1^2\OY_2\OY_j\big](\omu_2-\mu_2)(\omu_i-\mu_i)\\
\quad+\diy E\big[\OY_1^2\OY_i\OY_j\big](\omu_2-\mu_2)^2+E\big[\OY_1^2\big](\omu_2-\mu_2)^2(\omu_i-\mu_i)(\omu_j-\mu_j)\\
\quad+2\diy\;E\big[\OY_1\OY_2^2\OY_i\big](\omu_1-\mu_1)(\omu_j-\mu_j)+2\;E\big[\OY_1\OY_2^2\OY_j\big](\omu_1-\mu_1)(\omu_i-\mu_i)\\
\quad+4\diy\;E\big[\OY_1\OY_2\OY_i\OY_j\big](\omu_1-\mu_1)(\omu_2-\mu_2)+
4\;E\big[\OY_1\OY_2\big]\prod\limits_{k=1}^2(\omu_k-\mu_k)(\omu_i-\mu_i)(\omu_j-\mu_j)\\
\quad+2\diy\;E\big[\OY_1\OY_i\big](\omu_j-\mu_j)(\omu_1-\mu_1)(\omu_2-\mu_2)^2+2\;E\big[\OY_1\OY_j\big](\omu_i-\mu_i)(\omu_1-\mu_1)(\omu_2-\mu_2)^2\\
\quad+E\big[\OY_2^2\OY_i\OY_j\big](\omu_1-\mu_1)^2+E\big[\OY_2^2\big](\omu_1-\mu_1)^2(\omu_i-\mu_i)(\omu_j-\mu_j)\\
\quad+2\diy\;E\big[\OY_2\OY_i\big](\omu_2-\mu_2)(\omu_1-\mu_1)^2(\omu_j-\mu_j)+2\;E\big[\OY_2\OY_j\big](\omu_2-\mu_2)(\omu_1-\mu_1)^2(\omu_i-\mu_i)\\
\quad+E\big[\OY_i\OY_j\big]\prod\limits_{k=1}^2(\omu_k-\mu_k)^2+\prod\limits_{k=1}^2(\omu_k-\mu_k)^2 (\omu_i-\mu_i)(\omu_j-\mu_j).\end{array}\eeq
Follow the expectations from (\ref{Ap3}). Hence the final relation is concluded. $\quad$ $\Box$

\newpage

{\bf $\overline{\Lambda}_{ij}$ and $\Upsilon_{ij}$ in Example \ref{exam2}:}
\vskip .5 truecm
\beqq \begin{array}{l}
\overline{\Lambda}_{11}=\diy 12\rho^5(2-\rho)+3\rho^3(2-\rho)^3+4x_3^2\rho^2(2-\rho)^2(1-\rho)^2\\
\qquad+2\diy \rho^4 x_3^2(1-\rho)^2+x_3^4\rho(2-\rho)(1-\rho)^4-12\rho^3x_3^2(2-\rho)(1-\rho)^2,\\
\overline{\Lambda}_{12}=\overline{\Lambda}_{21}=-9\rho^4(2-\rho)^2-6\rho^6-6x_3^2(1-\rho)^2\rho^3(2-\rho)-\rho^2x_3^4(1-\rho)^4\\
\qquad +8\diy\rho^4x_3^2(1-\rho)^2+4\rho^2x_3^2(1-\rho)^2(2-\rho)^2,\\
\overline{\Lambda}_{22}=3\rho^4(2-\rho)^3+12\rho^5(2-\rho)+4\rho^2x_3^2(2-\rho)^2(1-\rho)^2+2\rho^4x_3^2(1-\rho)^2\\
\qquad+\diy x_3^4\rho(2-\rho)(1-\rho)^4-12 x_3^2\rho^3(2-\rho)(1-\rho)^2.
\end{array}\eeqq
Furthermore
\beqq \begin{array}{l}
\Upsilon_{11}=\diy 12\rho^5(2-\rho)+3\rho^3(2-\rho)^3+2(1-\rho)^2x_3^2\big(\rho^2(2-\rho)^2+2\rho^4\big)\\
\quad\diy -24(1-\rho)^2 x_3^2\rho^3(2-\rho)+3(1-\rho)^2x_3^2\rho^2(2-\rho)^2+(1-\rho)^6x_3^6\\
\quad\diy +4(1-\rho)^2x_3^2\big(\rho^2(2-\rho)^2-2\rho^3(2-\rho)\big)-8\rho^2(1-\rho)^4x_3^4+7\rho(2-\rho)(1-\rho)^4x_3^4.\end{array}\eeqq

\beqq\begin{array}{l}
\Upsilon_{12}=\Upsilon_{21}=-9\rho^4(2-\rho)^2-6\rho^6+9(1-\rho)^2x_3^2\big(\rho^2(2-\rho)^2+2\rho^4\big)+(1-\rho)^6x_3^6\\
\quad \diy -18(1-\rho)^2x_3^2 \rho^3(2-\rho)+6\rho(2-\rho)(1-\rho)^4x_3^4-6\rho^2(1-\rho)^4x_3^4. \end{array}\eeqq

\beqq \begin{array}{l}
\Upsilon_{22}=\diy 3\rho^4(2-\rho)^3+12\rho^5(2-\rho)+6(1-\rho)^2 x_3^2\big(\rho^2(2-\rho)^2+2\rho^4\big)\\
\quad \diy -24(1-\rho)^2 x_3^2\rho^3(2-\rho)+(1-\rho)^6x_3^6+7\rho (2-\rho)(1-\rho)^4x_3^4\\
\quad \diy -8\rho^2(1-\rho)^4x_3^4+ 3 \rho^2(2-\rho)^2(1-\rho)^2 x_3^2.\end{array}\eeqq

\vskip .5 truecm
{\emph{Acknowledgements --}}
SYS thanks the CAPES PNPD-UFSCAR Foundation for the financial support in the year 2014-5. SYS thanks
the Federal University of Sao Carlos, Department of Statistics, for hospitality during the year 2014-5.





\end{document}